\documentclass[twocolumn,showpacs,preprintnumbers,amsmath,amssymb,prd]{revtex4}
\usepackage{graphicx,psfrag}
\usepackage{dcolumn}
\usepackage{bm}

\newcommand{\ie}{\textit{i.e.}}
\newcommand{\Nc}{N_{\rm c}}
\newcommand{\Nf}{N_{\rm f}}

\begin{document}

\preprint{}

\title{Possible Ferromagnetism in the Large $\Nc$ and $\Nf$ limit of quark matter}

\author{Kazuaki Ohnishi}
 \email{kohnishi@th.phys.titech.ac.jp}
\author{Makoto Oka}
 \email{oka@th.phys.titech.ac.jp}
\author{Shigehiro Yasui}
 \email{yasui@th.phys.titech.ac.jp}
\affiliation{Department of Physics, H-27, Tokyo Institute of Technology,
Meguro, Tokyo 152-8551, Japan}
\date{\today}

\begin{abstract}
We consider high density quark matter in the large $\Nc$ and $\Nf$ limit with
$\Nf/\Nc$ fixed. In this limit, the color superconductivity disappears. We
discuss that the chiral density wave state is also absent in the limit, if we
assume the existence of the non-perturbative magnetic screening effect as
indicated by recent lattice study. We argue that ferromagnetism can become
a candidate for the ground state if quarks are massive.
\end{abstract}


\maketitle

There have been extensive studies of high density quark matter
\cite{Rajagopal:2000wf}, which may
be realized in the core of neutron stars and may be accessible in the
relativistic heavy ion collision experiments. Due to the asymptotic freedom
of QCD, the interaction between quarks at high density is dominated
by a one-gluon-exchange (OGE) process. In the OGE interaction we have an
attractive channel (the color anti-symmetric $\underline{3}$ channel), which
inevitably induces the Cooper instability near the Fermi surface. Thus the
ground state of quark matter at high density region would be
in a BCS state, \ie, the color superconductivity
\cite{Barrois:1977xd,Bailin:1983bm}.

In the high density quark matter, three light flavors are relevant, \ie, up,
down and strange quarks, which have small but non-zero masses
\cite{Itoh:1970uw}. In the SU(3) chiral limit, the ground state is supposed
to be the Color-Flavor-Locked (CFL) phase \cite{Alford:1998mk}. Indeed,
because all the three colors can participate in the Cooper pairing, the CFL
phase is favored compared to the 2SC state \cite{Alford:1997zt,Rapp:1997zu},
where only two colors are in the superconducting state.
In the CFL state, the three flavors are locked to the three colors and
consequently the chiral symmetry is spontaneously broken according to
${\rm SU}_{\rm color}(3)\otimes{\rm SU}_{\rm L}(3)\otimes
{\rm SU}_{\rm R}(3)\rightarrow {\rm SU}_{\rm color+L+R}(3)$.
Thus the superconducting state at the high enough density, where the quark
masses of the three flavors can be neglected compared with the chemical
potential, is believed to be in the CFL phase.

Although we have three colors in this world, it would be fruitful to consider
an SU($\Nc$) gauge theory treating the number of color, $\Nc$, as a free
parameter \cite{'tHooft:1973jz,Witten:1979kh,Manohar:1998xv}.
In fact, a meaningful large $\Nc$ limit is obtained by making the gauge coupling
$g$ scale as $g\sim 1/\sqrt{\Nc}$ and by carrying out an expansion in terms of
$1/\Nc$ around the large $\Nc$ limit.

In the large $\Nc$ expansion, the leading term is given by planar diagrams,
while non-planar diagrams contribute to the next-to-next leading term of
$\mathcal{O}(1/\Nc{}^{2})$. The diagrams with one quark loop are suppressed
by a $1/\Nc$ factor, giving the next-to-leading contribution.

It would be of natural interest to ask what happens to high density quark
matter if we take the large $\Nc$ limit. The color
superconductivity vanishes in the large $\Nc$ limit because
the Cooper pair is not a color singlet
\cite{Deryagin:1992rw,Shuster:1999tn,Park:1999bz,Rapp:2000zd,
Zhitnitsky:2005nb,McLerran:2007qj}. Actually, the superconducting gap $\Delta$
in the perturbative regime is estimated to be \cite{Son:1998uk}
$\Delta\sim\mu\exp(-\sqrt{6\Nc/(\Nc+1)}\pi^{2}/g)$, which goes to zero
exponentially as $\Nc\rightarrow\infty$. This means that the color
superconductivity is a phenomenon that can never seen in the $1/\Nc$ expansion,
that is, a non-perturbative phenomenon with respect to the $1/\Nc$
expansion, like the pair creation process of baryon-anti-baryon through a
virtual photon \cite{Witten:1979kh}.

Instead of the color superconductor, the ground state of quark matter in the
large $\Nc$ limit is believed to be replaced by the chiral density wave state
\cite{Deryagin:1992rw,Shuster:1999tn,Park:1999bz,Rapp:2000zd}, in which
the condensate of particle-hole pairs is deformed with a finite spatial
wave number.

We have another free parameter in QCD, \ie, the number of flavor, $\Nf$.
It is meaningful to take the large $\Nc$ and $\Nf$ limit simultaneously with
$\Nf/\Nc$ fixed \cite{Veneziano:1976wm,Montanet:1980te,Luty:1994ua}.
In this limit, the asymptotic freedom of QCD is preserved
\cite{Manohar:1998xv}.
One of the features of the large $\Nc$ and $\Nf$ limit is that quark loops
are not suppressed. This is because the factor $\Nf$ arising from the quark
loop compensates the suppression factor $1/\Nc$. Thus all the planar diagrams
including quark loops constitute the leading contribution in the expansion.
We also note that the ${\rm U}_{\rm A}(1)$ anomaly is not suppressed in the
limit, which is of $\mathcal{O}(\Nf/\Nc)$ \cite{Luty:1994ua}.

The question we wish to address here is what is the ground state of high
density quark matter in the large $\Nc$ and $\Nf$ limit. The question
would be reasonable because for the CFL phase, the same number of flavor as
that of color is essential: $\Nf=3$ is as large as $\Nc=3$. The large $\Nc$
and $\Nf$ limit would be an appropriate competitor to the CFL phase. In the
limit, the color superconductivity disappears for the reason as mentioned
above \cite{Schafer:1999fe}. The chiral density wave state can also be absent
in the limit. This can be seen as follows.
In the three spatial dimension,
the chiral density wave is induced by the infra-red singularity stemming
from the long range gluon interaction \cite{Shuster:1999tn}.
In quark matter, generally speaking, the infinite range interaction can be
cut off by screening due to the quark loops. Actually, the electric gluon
is screened perturbatively, which acquires the Debye mass of
$M_{\rm ele}{}^{2}\sim\Nf g^2 \mu^2$ \cite{Kapusta}. As for the magnetic
sector, it is known that the screening does not appear within perturbation
theory \cite{Son:1998uk}. However, there is still a possibility that
the magnetic screening is generated by some non-perturbative mechanism.
In fact, recent lattice study shows that we can have the magnetic screening
at finite density \cite{Hands:2006ve}. In the following, let us adopt this
possibility.
In the large $\Nc$ but small $\Nf$ limit, the quark
loops are suppressed and these screening effects disappear, resulting in
the instability of the chiral density wave.
In the large $\Nc$ and $\Nf$
limit, however, the quark loops are not suppressed as we mentioned.
This is explicitly seen by the electric mass
$M_{\rm ele}{}^{2}\sim\Nf g^2 \mu^2$ which remains finite in the limit.
Thus the chiral density wave state cannot occur and
we are faced with the problem: What is the ground state of
quark matter at the high density region in the large $\Nc$ and $\Nf$ limit.

In this report, we will not be able to give a decisive answer to the question.
We will only explore one possible candidate.

Apart from the possibility that quark matter is in a normal state in the limit,
let us suppose the existence of some condensate breaking some symmetry. In order
for the condensate to survive when we take $\Nc$ to infinity, it needs to be a
color singlet object in the form of $\langle\bar{q}\Gamma q\rangle$ involving
some gamma matrices $\Gamma$. Since in the large $\Nc$ and $\Nf$ limit, spin
singlet condensate is not available, the candidate state that should be
considered in the first place would be the state with the spin 1 condensate
$\langle\bar{q}\gamma_{\mu}\gamma_{5} q\rangle$, that is, the ferromagnetism
(FM)
\cite{Fantoni:2001ih,Vidana:2002pc,Tatsumi:1999ab,Maruyama:2000cw,
Nakano:2003rd,Tatsumi:2003bk,Tatsumi:2005zh,Tatsumi:2005ri,Niegawa:2003bd,
Niegawa:2004vd,Maedan:2006ib}.

FM in dense hadronic and quark matter has been studied by several authors.
One of their motivations is to explain the observed strong magnetic field in
compact stars \cite{Chanmugam:1992rz,Anderson}. The theoretical possibility
of FM in quark matter was first argued by Tatsumi \cite{Tatsumi:1999ab}.
For the non-relativistic itinerant electron gas, it was suggested
by Bloch \cite{Bloch} that FM can appear as a consequence of competition
between the kinetic energy and the Coulomb potential energy. Tatsumi extended
the argument to quark matter to show that the OGE interaction between quarks
can induce the instability for FM in both the ultra-relativistic and
non-relativistic regimes with somewhat different mechanisms. It would be
worth while to examine FM in quark matter to see whether or not it can survive
in the large $\Nc$ and $\Nf$ limit. It is noted that FM should persist in the
large $\Nc$ limit because the condensate $\langle\bar{q}\gamma_{\mu}\gamma_{5}
q\rangle$ is a color singlet. We are concerned with how FM depends on $\Nf$.
For this purpose, let us recapitulate Tatsumi's argument using the OGE
approximation with special attention to the $\Nf$ (and $\Nc$) dependence.
In the end, we will see that FM can become a candidate of the ground state in
the limit if quarks are massive.

In the Landau theory for the weakly interacting Fermi-liquid, the total energy
density is given by \cite{Baym:1975va}
\begin{align*}
\epsilon_{\rm total} =&\epsilon_{\rm kinetic}+\epsilon_{\rm potential}
\\
=& \sum_{\sigma}\int\frac{{\rm d}^3p}{(2\pi)^3}E_{p}n_{\boldsymbol p}
\\
&+\frac{1}{2}\sum_{\sigma\sigma'}\int\frac{{\rm d}^3p}{(2\pi)^3}
\int\frac{{\rm d}^3p'}{(2\pi)^3}
f_{{\boldsymbol p}\sigma, {\boldsymbol p'}\sigma'}
n_{\boldsymbol p}n_{\boldsymbol p'},
\end{align*}
where the Landau Fermi-liquid interaction
$f_{{\boldsymbol p}\sigma, {\boldsymbol p'}\sigma'}$ is related to the
two-particle forward scattering amplitude
$\mathcal{M}_{{\boldsymbol p}\sigma, {\boldsymbol p'}\sigma'}$ as
\begin{equation*}
f_{{\boldsymbol p}\sigma, {\boldsymbol p'}\sigma'}
=\frac{m}{E_{p}}\frac{m}{E_{p'}}
\mathcal{M}_{{\boldsymbol p}\sigma, {\boldsymbol p'}\sigma'}.
\end{equation*}
We have used obvious notations: $m$ is the quark mass and
$E_{p}=\sqrt{{\boldsymbol p}^2+m^2}$. $\sigma$ stands for the spin degree
of freedom.

For the kinetic energy density, there arises an overall factor of $\Nc\Nf$
because we have $\Nc\Nf$ Fermi spheres. The potential energy density would
consist of two terms associated with the direct and exchange scattering
processes, respectively. However, the former vanishes because it involves
${\rm tr} \lambda_{a}$ in the OGE approximation. In the remaining exchange
process, scattering of two quarks with different flavors can not contribute
(\ie, the processes such as $ud\rightarrow du$ are forbidden) because the
OGE interaction does not mix flavors
\footnote{For finite $\Nc$ and $\Nf$, we have finite instanton effects
\cite{Kiuchi:2005xu}. It would be interesting to consider the instanton effect
which mixes flavors.}.
Only the processes involving the
identical flavor such as $uu\rightarrow uu$ are allowed. This means that
the Fermi sphere of each flavor makes an independent contribution. Thus
the potential energy density receives a factor $\Nf$. On the other hand,
the quarks with different colors can take part in the exchange process,
giving rise to a factor $\Nc{}^2$. Eventually, the potential energy density is
proportional to $\Nf\Nc{}^2g^2$, which is the same order as the kinetic energy
density. Thus, the factor $\Nc\Nf$ factorizes out of the total energy density
and the competition between the kinetic and potential energies is not influenced
by the numbers of color and flavor. If FM appears at some arbitrary numbers
of color and flavor, it will persist in the large $\Nc$ and $\Nf$ limit.
The large number of flavor neither encourages nor discourages FM.

Now we consider the higher order diagrams beyond the OGE approximation.
Actually, in Refs.\ \cite{Nakano:2003rd} and \cite{Niegawa:2004vd}, attempts
were made to resum some kind of the infinite number of diagrams.
Unfortunately, in $(3+1)$ dimension, it is not possible to resum
all the leading order diagrams in the large $\Nc$ expansion, in contrast to
the $(1+1)$ dimension where the Hartree-Fock approximation gives the exact
solution in the large $\Nc$ limit \cite{Witten:1979kh}. In fact,
in Ref.\ \cite{Nakano:2003rd}, the analysis was performed in the Hartree-Fock
approximation based on the one-gluon-exchange scattering, that is, the
so-called ladder QCD, which is not complete in view of the large $\Nc$ and
$\Nf$ expansion. There exist three types of diagrams that give rise
to the leading contributions in the large $\Nc$ and $\Nf$ expansion, which
are missed in the ladder QCD analysis. First, in the ladder QCD,
interactions between gluons are ignored.
However, planar diagrams with gluonic interactions give a leading order
contribution. Second, we may consider the
multi-gluon exchange scattering process, which has the same topology as
the OGE process and thus can give the leading contribution in the expansion.
The Fock terms in the multi-gluon exchange scattering process should be
included. Third, in the multi-gluon exchange process, not only the Fock
but also the Hartree (or direct) terms survive because the trace over
color does not vanish generally. The Hartree terms are of the leading order
in the large $\Nc$ and $\Nf$ expansion.
This counting can be understood as follows.
If we consider the self-energy diagrams
for the Hartree and Fock contributions, we find that the Hartree diagrams
involve one quark loop. Thus the Hartree terms are next to leading order
in the large $\Nc$ expansion.
However, $\Nf$ coming from the quark loop
makes the Hartree terms of the same order as the Fock term. Thus, in the
large $\Nc$ and $\Nf$ expansion (but not in the large $\Nc$ expansion),
the Hartree term can become the leading contribution.

It is ideal that all the above diagrams are considered in order to
investigate FM in the $\Nc$ and $\Nf$ limit. In particular, the Hartree
terms in the multi-gluon exchange would be important, which
might tend to disfavor FM, contrary to the Fock terms. However,
the diagrammatic analysis of the whole leading contributions is a formidable
task. It would be necessary to use other non-perturbative methods such as
the lattice QCD and the AdS/QCD
\cite{Maldacena:1997re,Sakai:2004cn,Kim:2006gp,Horigome:2006xu,
Nakamura:2006xk,Kobayashi:2006sb}, although both the methods involve some
difficulty at present; the former suffers from the so-called sign problem
and the latter is restricted to the $\Nf \ll \Nc$ case only. In this
situation, it would not be meaningless to discuss the results derived in
the analyses in Refs.\ \cite{Nakano:2003rd} and \cite{Niegawa:2004vd},
which partly take account of the leading order diagrams in the $\Nc$ and
$\Nf$ expansion. Especially, we believe that the Hartree terms do not
overwhelm the Fock terms because the former is a contribution
beyond the OGE diagram.
Thus, at least the qualitative features
in Refs.\ \cite{Nakano:2003rd} and \cite{Niegawa:2004vd} would hold.

We repeat the results of these two works, which will be our conclusion in
this report.

\begin{itemize}
\item If quarks are massless, it can be analytically proven that FM does not
 appear \cite{Nakano}. This is because the OGE conserves chirality
and thus the helicity of massless quarks, and the Fermi seas for the
right-handed ({\it i.e.} positive helicity) and left-handed
(negative helicity) quarks become the same. Thus, in this case, FM can not
be a candidate of the ground state of quark matter at high density
in the large $\Nc$ and $\Nf$ limit.

\item When quarks have finite masses, FM can appear. The large quark mass
is favorable for developing the ferromagnetic condensate.
The density region where FM shows up depends on the interaction range between
quarks: If we assume the range to be infinite \cite{Niegawa:2004vd}, FM
emerges in the lower density region below some critical density. On the other
hand, if we adopt the zero range approximation \cite{Nakano:2003rd}, FM is
developed in the higher density region including the asymptotically high
density above a critical density.
The higher the density is, the larger the spin polarization arises.
This approximation may be justified if we take account of the
finite screening mass, which survives in the large $\Nc$ and $\Nf$ limit.
In reality, the range will be non-zero finite even with the screening effect
and thus a careful examination is necessary to determine which limiting
situation is more appropriate.
In either case, for massive quarks, FM can become a candidate of the ground
state of quark matter in the large $\Nc$ and $\Nf$ limit.
\end{itemize}

Before summary, two comments are in order.

First, in the strong coupling regime which may be relevant to the intermediate
density region, the chiral density wave can emerge irrespective of the
interaction range \cite{Rapp:2000zd}. The ground state in the large $\Nc$ and
$\Nf$ limit would
be the chiral density wave state. Interplaying with the chiral density wave,
the spin density wave associated with FM may appear
\cite{Tatsumi:2004dx,Nakano:2004cd,Tatsumi:2005zh}.

Second, it does not seem that $\Nc=\Nf=3$ is so large that FM becomes the
ground state to compete with CFL in the real world. However $\Nc=\Nf=3$ might
be large enough that FM gives a prominent meta-stable state relevant to
compact stars \cite{Tatsumi:1999ab}. If this is the case, the strong  magnetic
field in compact stars could be regarded as a remnant of FM prevailing
in the large $\Nc$ and $\Nf$ limit.

To summarize, we have addressed the question of what is the ground state
of high density quark matter in the large $\Nc$ and $\Nf$ limit.
If we assume the non-perturbative magnetic screening as suggested by
lattice calculation, the chiral density wave state as well as the color
superconductivity are absent. We have proposed the ferromagnetism (FM) as
a candidate. It would be interesting to note that even if there is no
magnetic screening, FM and the chiral density wave might compete with each
other because the latter is weakened which is now induced only by the
magnetic gluon and to which the electric gluon does not contribute
\footnote{As discussed in summary of Ref.\ \cite{Deryagin:1992rw},
within the ladder approximation, the electric and magnetic gluons contribute
to the chiral density wave independently so that only the coefficient of
the gap equation is modified when the electric gluon is screened and
inactive.}.
We have seen that FM can remain in the limit within the OGE approximation. 
We have discussed the leading diagrams in the large $\Nc$ and $\Nf$ expansion
beyond OGE to find that there exist contributions that are not taken care of
within the ladder QCD. It will be necessary to study further the large $\Nc$ and
$\Nf$ limit of FM.
Based on the works of Refs.\ \cite{Nakano:2003rd} and \cite{Niegawa:2004vd},
which we believe are reliable at least qualitatively,
we conclude that FM can become a candidate if quarks are massive.
FM may appear below or above a critical density, depending on the
interaction range between quarks. In the massless case, FM is totally
absent. We are left with an open question for this idealized world.
The normal state would be one of the candidates.
We note that even in the massive case, there still remains
possibility that another state gives the true ground
state with lower free energy. Drawing the phase diagram in the
$(\Nc, \Nf)$ plane will be a theoretical challenge
\cite{Blaizot:2005fd,Frandsen:2005mb}.

\begin{acknowledgments}
The authors are grateful to Dr.\ Eiji Nakano for useful discussion.
One of the authors (K.\ O.\ ) thanks Dr.\ Shin Nakamura for useful
communication.
This work is supported by Grants-in-Aid for Scientific Research in
Priority Areas 1707002, and (B) 15340072.
\end{acknowledgments}


\end{document}